\documentclass[aps,prb,twocolumn,10pt,superscriptaddress,notitlepage]{revtex4-1}

\newcommand{\papertitle}{Electrical detection of hyperbolic phonon-polaritons in heterostructures of graphene and boron nitride}

\usepackage[utf8]{inputenc}
\usepackage{graphicx}

\usepackage{graphicx}
\usepackage{amsmath}
\usepackage{natbib}
\usepackage{float}
\usepackage{siunitx}
\usepackage{lmodern}
\usepackage[T1]{fontenc}
\usepackage[unicode=true,
	    colorlinks=true,
	    linkcolor=blue,
	    citecolor=blue,
	    urlcolor=blue]{hyperref} 
\usepackage{sansmath,caption}
\newcommand{\newtextbar}[1][3]{\scalebox{#1}[1]{\textbar}}
\DeclareCaptionLabelSeparator{vertline}{\textbf{~\newtextbar}}
\usepackage[font=sf, labelfont={sf,bf},size=small,justification=RaggedRight,textfont={sf,sansmath},labelsep=vertline]{caption}

\renewcommand{\Re}{\operatorname{Re}}
\renewcommand{\Im}{\operatorname{Im}}

\begin{document}
\title{\Large\textsf{\papertitle}}

\author{Achim Woessner}
\thanks{These authors contributed equally}
\affiliation{ICFO – Institut de Ciències Fotòniques, The Barcelona Institute of Science and Technology, 08860 Barcelona, Spain}
\author{Romain Parret}
\thanks{These authors contributed equally}
\affiliation{ICFO – Institut de Ciències Fotòniques, The Barcelona Institute of Science and Technology, 08860 Barcelona, Spain}
\author{Diana Davydovskaya}
\affiliation{ICFO – Institut de Ciències Fotòniques, The Barcelona Institute of Science and Technology, 08860 Barcelona, Spain}
\author{Yuanda Gao}
\affiliation{Department of Mechanical Engineering, Columbia University, New York, NY 10027, USA}
\author{Jhih-Sheng Wu}
\affiliation{University of California San Diego Department of Physics 0319 9500 Gilman Drive La Jolla, CA 92093-0319}
\author{Mark B. Lundeberg}
\affiliation{ICFO – Institut de Ciències Fotòniques, The Barcelona Institute of Science and Technology, 08860 Barcelona, Spain}
\author{Sébastien Nanot}
\affiliation{ICFO – Institut de Ciències Fotòniques, The Barcelona Institute of Science and Technology, 08860 Barcelona, Spain}
\author{Pablo Alonso-González}
\affiliation{\footnotesize CIC nanoGUNE, 20018 Donostia-San Sebastian, Spain}
\affiliation{\footnotesize Departamento de Física, Universidad de Oviedo, 33007, Oviedo, Spain}
\author{Kenji Watanabe}
\affiliation{National Institute for Materials Science, 1-1 Namiki, Tsukuba 305-0044, Japan}
\author{Takashi Taniguchi}
\affiliation{National Institute for Materials Science, 1-1 Namiki, Tsukuba 305-0044, Japan}
\author{Rainer Hillenbrand}
\affiliation{\footnotesize CIC nanoGUNE and UPV/EHU, 20018 Donostia-San Sebastian, Spain}
\affiliation{\footnotesize IKERBASQUE, Basque Foundation for Science, 48011 Bilbao, Spain}
\author{Michael M. Fogler}
\affiliation{University of California San Diego Department of Physics 0319 9500 Gilman Drive La Jolla, CA 92093-0319}
\author{James Hone}
\affiliation{Department of Mechanical Engineering, Columbia University, New York, NY 10027, USA}
\author{Frank H.L. Koppens}
\email{frank.koppens@icfo.eu}
\affiliation{ICFO – Institut de Ciències Fotòniques, The Barcelona Institute of Science and Technology, 08860 Barcelona, Spain}
\affiliation{ICREA – Institució Catalana de Recerça i Estudis Avancats, 08010 Barcelona, Spain}


\maketitle

\begin{figure*}[t!]
\centering
\includegraphics{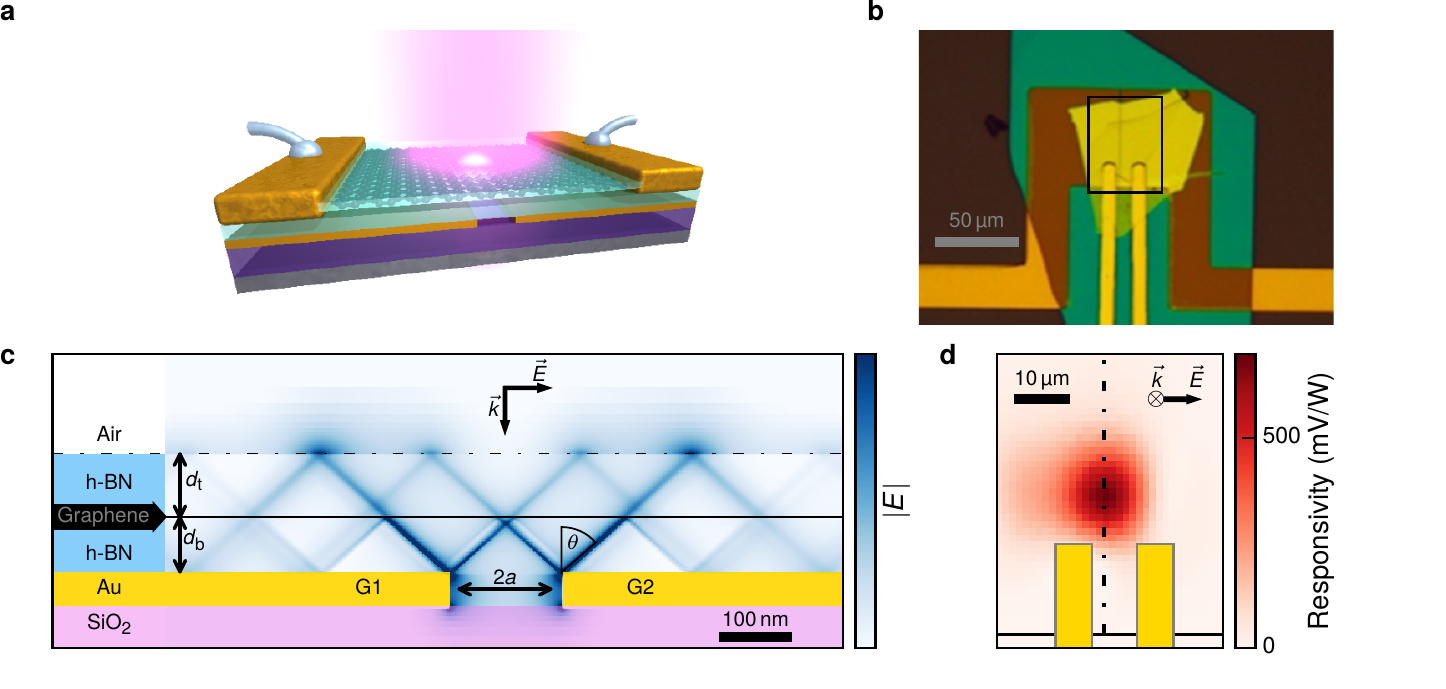}
\caption{
\label{fig1}
\textbf{Device schematic and working principle.}
\textbf{a} Schematic of the encapsulated graphene pn-junction. 
The h-BN/graphene/h-BN is placed on two gold split gates and contacted electrically by the edges. 
\textbf{b} Optical image of one device: the top (bottom) BN is blue (yellow).
The black rectangle indicates the measurement region in \textbf{d}.
The distance between the two gates in this device is 100\,nm.
\textbf{c} Side view of the propagating hyperbolic phonon-polaritons simulated by FDTD for the device in \textbf{b}.
The thickness of the bottom h-BN is 50\,nm and the top h-BN is 55\,nm.
HPPs are launched at the edges of the split gates and propagate as directional rays.
While they cross the graphene plane they are partially absorbed leading to a temperature increase in the graphene. 
\textbf{d} Spatial map of the device responsivity for a polarization of the laser perpendicular to the gap of the split gate.
The photoresponse arises at the junction, indicated by the dashed-dotted line.
The graphene edge is indicated by the solid black line.
The gate voltages used here are ($V_\mathrm{g1} = 1.2$\,V and $V_\mathrm{g2} = -0.21$\,V).
The electric field polarization and propagation direction ($E$,$k$) are represented in panels \textbf{c} and \textbf{d}.
}
\end{figure*}

\begin{figure*}[t]
\centering
\includegraphics{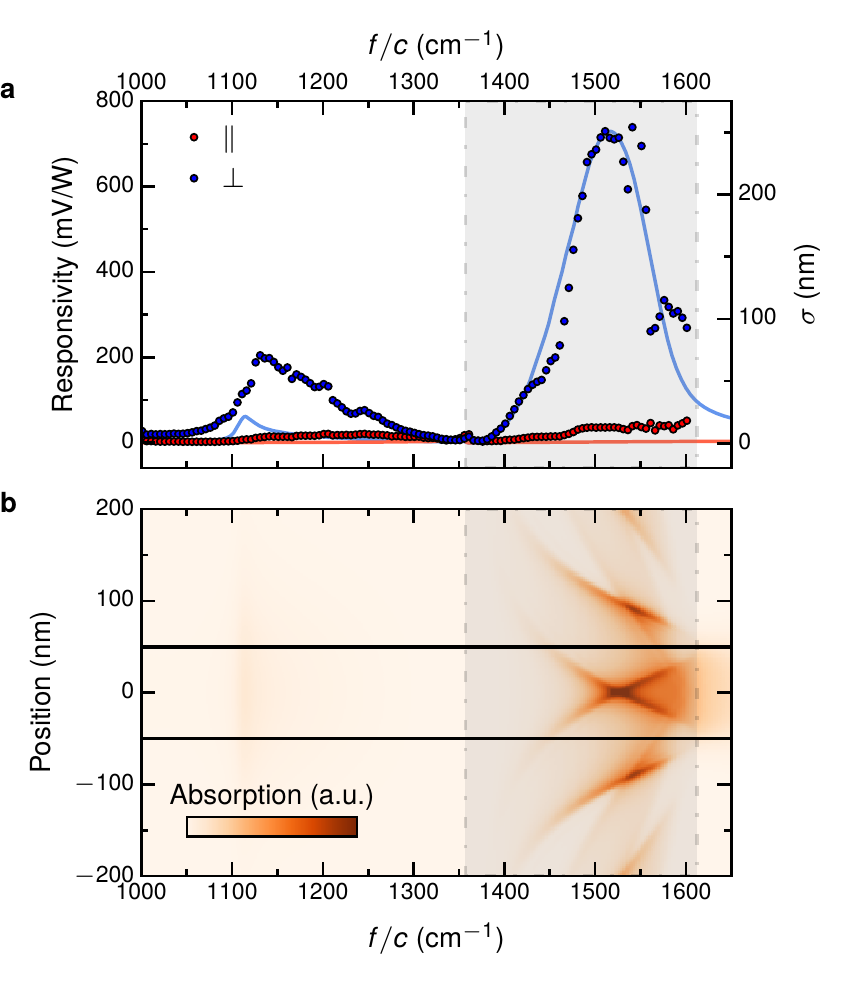} %
\includegraphics{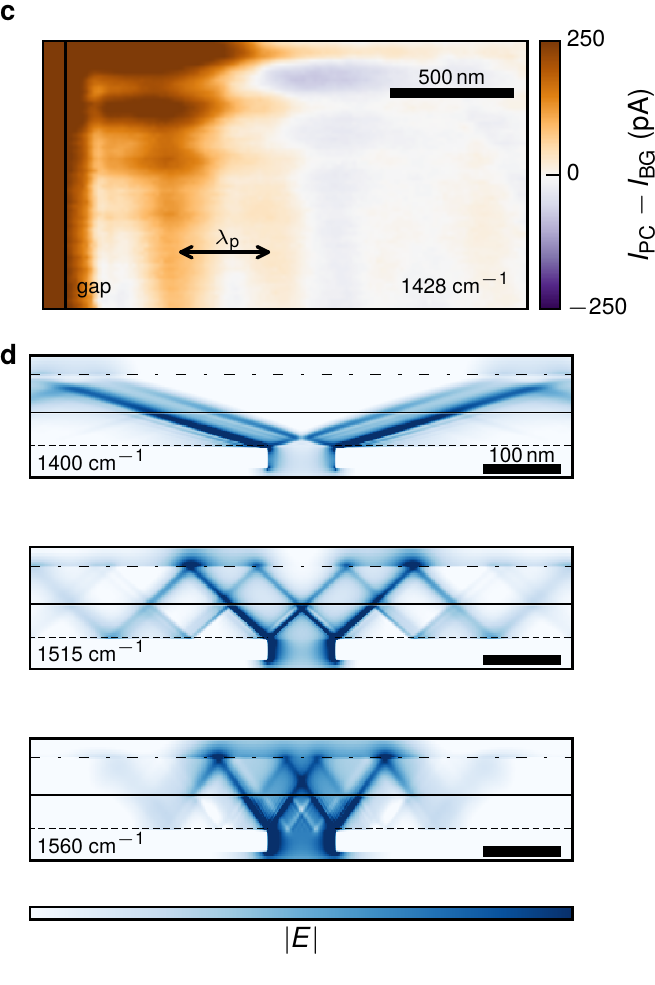}

\caption{
\label{fig2}
\textbf{Absorption and photocurrent spectra.}
\textbf{a} Responsivity spectrum for light polarisation perpendicular (parallel) to the junction in blue (red).
The gate voltages used here are (Vg1 = 1.2 V and Vg2 = -0.21 V). 
The main peak which lies in the upper reststrahlen band of h-BN (grey shaded region) is the result of the hyperbolic phonon-polariton assisted photoresponse.
Solid lines are absorption spectra simulated by FDTD. 
\textbf{b} FDTD simulation of the spatial absorption profile in the vicinity of the junction as a function of the laser frequency.
The spatial integral at each frequency is proportional to the simulated absorption cross section spectrum in \textbf{a}.
\textbf{c} Near-field photocurrent measurement with subtracted background photocurrent of a similar device with 42\,nm of bottom h-BN thickness, 13\,nm on top and a gap width of 50\,nm.
The top edge of the panel is the edge of the graphene and the gap position is indicated.
The left gate is set to $-2$\,V and the right gate to $0.1$\,V.
\textbf{d} Side views of the propagating HPPs at 1400\,cm$^{-1}$ (below the maximum), 1515\,cm$^{-1}$ (at the maximum) and at 1560\,cm$^{-1}$ (above the maximum) respectively.  
}
\end{figure*}

\begin{figure*}[t]
\centering
\includegraphics{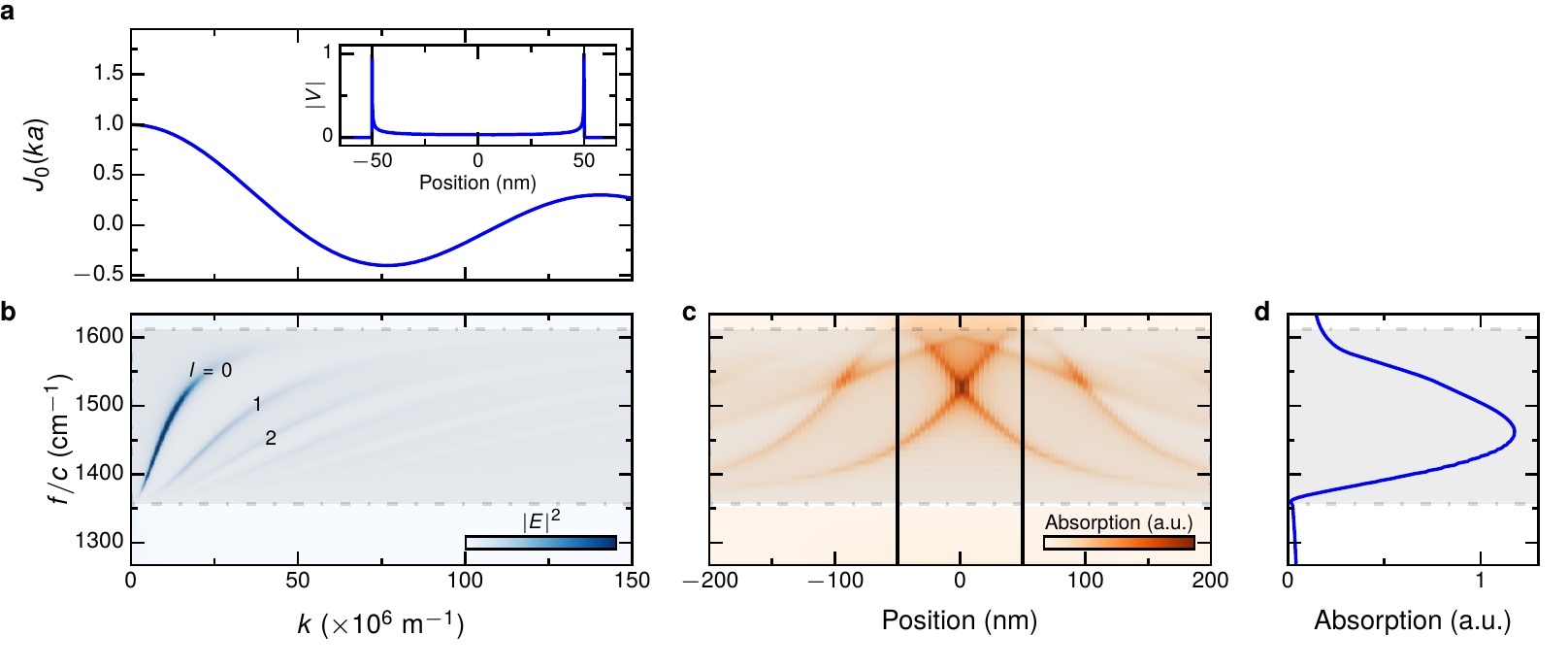} %
\caption{
\label{fig3}
\textbf{Analytic calculation of absorption spectra.}
\textbf{a} Momentum distribution provided by a metallic split gate with 100\,nm gap width.
The associated electric field profile is shown in the inset. 
\textbf{b} HPP frequency dispersion curves showing discrete eigenmodes $l = 0, 1,2,...$ calculated for the full system of a metal gate, 50\,nm h-BN, graphene and 55\,nm h-BN on top.  
\textbf{c} Real space absorption profile obtained from the analytic calculations.
\textbf{d} The resulting absorption spectrum calculated in $k$-space of the total power absorbed inside the graphene.
}
\end{figure*}

\begin{figure}[t]
\centering
\includegraphics{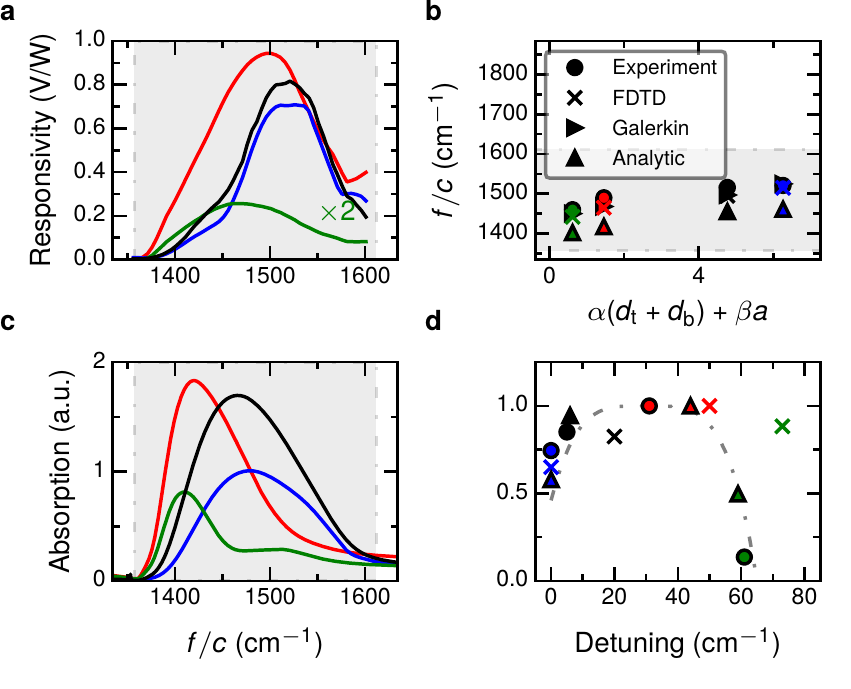} %
\caption{
\label{fig4}
\textbf{Comparison between experiments, simulation and analytic model.}
\textbf{a} Responsivity spectra of the investigated devices.
\textbf{b} Comparison of the peak frequency of the devices obtained experimentally, by FDTD simulation, by the Galerkin method (see supplement) and using the analytic model. 
Where $d_\mathrm{t}$ and $d_\mathrm{b}$ are the top and bottom thicknesses of the h-BN respectively and $a$ is the gap width. 
The parameters $\alpha$ and $\beta$ are extracted from the analytic model (see supplement). 
\textbf{c} Absorption spectra calculated analytically. 
\textbf{d} Normalized peak values as a function of the detuning (see text) of its frequency. 
The gray dashed-dotted line is a guide to the eye.
}
\end{figure}

\textsf{\textbf{
Light properties in the mid-infrared can be controlled at a deep subwavelength scale using hyperbolic phonons-polaritons (HPPs) of hexagonal boron nitride (h-BN).\cite{Caldwell2014,Dai2014b,Dai2015,Li2015,Giles2016,Basov2016a,Low2016a}
While propagating as waveguided modes\cite{Dai2015,Li2015,Yoxall2015} HPPs can concentrate the electric field in a chosen nano-volume.\cite{Caldwell2014,Dai2015} 
Such a behavior is at the heart of many applications including subdiffraction imaging\cite{Caldwell2014,Dai2015} and sensing.
Here, we employ HPPs in heterostructures of h-BN and graphene as new nano-optoelectronic platform by uniting the benefits of efficient hot-carrier photoconversion in graphene and the hyperbolic nature of h-BN.
We demonstrate electrical detection of HPPs by guiding them towards a graphene pn-junction.
We shine a laser beam onto a gap in metal gates underneath the heterostructure, where the light is converted into HPPs.
The HPPs then propagate as confined rays heating up the graphene leading to a strong photocurrent.
This concept is exploited to boost the external responsivity of mid-infrared photodetectors, overcoming the limitation of graphene pn-junction detectors due to their small active area and weak absorption.
Moreover this type of detector exhibits tunable frequency selectivity due to the HPPs, which combined with its high responsivity paves the way for efficient high-resolution mid-infrared imaging. 
}} 

Hexagonal boron nitride has found multiple uses in van der Waals heterostructures, such as a perfect substrate for graphene,\cite{Dean2010,Wang2013} a highly uniform tunnel barrier,\cite{Britnell2012,Withers2015} and an environmentally robust  protector.
In particular, h-BN substrates enable one to achieve high carrier mobility and homogeneity in graphene.\cite{Dean2010,Wang2013}
In addition, h-BN is a natural hyperbolic material as in the two so called reststrahlen bands (760--825\,cm$^{-1}$ and 1360--1610\,cm$^{-1}$) the in plane ($\epsilon_\mathrm{x,y}$) and the out of plane ($\epsilon_\mathrm{z}$) permittivity are of opposite sign.\cite{Caldwell2014,Dai2015,Li2015}
As a consequence, h-BN supports propagating hyperbolic phonon-polaritons (HPPs) which are electromagnetic modes\cite{Caldwell2014,Dai2015} originating in the coupling of photons to optical phonons. 
Because of their unique physical properties such as long lifetime, tunability,\cite{Dai2014b} slow propagation velocity,\cite{Yoxall2015} and strong field confinement the HPPs have a great potential for applications in nanophotonics.
The capability to concentrate light into small volumes can also have far-reaching implications for opto-electronic technologies, such as mid-infrared photodetection,\cite{Liang2011,Yan2012f,Freitag2013,Yao2014,Freitag2014,Sassi2016,Gopalan2017} on-chip spectroscopy and sensing.
These concepts, however, remain underexplored.

Here we present a hyperbolic opto-electronic device that takes taking advantage of that fact that h-BN is at the same time an ideal substrate for graphene as well as an excellent waveguide for HPPs.
We show how HPPs can be exploited to concentrate the electric field of incident mid-infrared beam towards a graphene pn-junction, where it is converted to a photovoltage.
The impact of the HPPs leads to a strongly increased responsivity of the graphene pn-junction in the mid-infrared up to 1\,V/W, with zero bias applied. 

In previous studies graphene pn-junctions have shown very high internal efficiencies\cite{Xu2010,Gabor2011,Lemme2011,Herring2014a,Hsu2015,Koppens2014} due to the strong photo-thermoelectric effect in graphene.
However, the active area of this type of devices is extremely small, leading to poor light collection.
For detecting mid-infrared light this issue is even more acute. 
By exciting hyperbolic phonon polaritons we strongly enhance the effective absorption.
We compare our experimental results with FDTD simulations and an analytical model, providing insight into the underlying physical processes and the frequency tunability of our novel mid-infrared detectors.

The investigated devices consist of heterostructures of monolayer graphene encapsulated in h-BN, obtained by the polymer-free van der Waals assembly technique,\cite{Wang2013} and placed on top of two metal gates separated by a narrow gap (Fig.~\ref{fig1}a). 
The graphene layer has a mobility of $\sim30000\,\mathrm{cm^2/Vs}$.
It is electrically connected to the source and drain electrodes by edge contacts\cite{Wang2013} (see methods).
An optical micrograph of a typical device is shown in Fig.~\ref{fig1}b.
The individually tunable carrier density on both sides of the split gate is used to tune the photosensitivity of our device.\cite{Gabor2011,Hsu2015,Lemme2011,Xu2010}

The operation of our device is as follows.
HPPs are launched at the sharp gold edges of the split gate when the laser beam illuminates the sample under normal incidence with the polarization perpendicular to the gap between the gate electrodes.\cite{Huber2008,Dai2015,Yoxall2015}
While the HPPs propagate as highly directional rays in both bottom and top h-BN slabs of the stack, they are absorbed when they pass through graphene (Fig.~\ref{fig1}c) creating hot carriers.
The hot carriers diffuse over a length scale of the electron cooling length (about 0.5--1\,$\si{\micro\metre}$) and generate a temperature increase peaking at the graphene junction defined by the position of the gap in the metal gates.
This inhomogeneous temperature distribution induces a photovoltage due to the Seebeck effect.
Thus, all HPPs absorbed within approximately one cooling length from the junction contribute to the photovoltage.
Similar HPPs lauching presumably also occurs at the source and drain gold contacts.
However, they do not contribute to the photovoltage because the electrodes are situated much further than the electron cooling length from the junction. 

The measured spatially resolved photoresponse of the device is shown in Fig.~\ref{fig1}d. 
The photoresponse arises mainly at the junction (shown as a dashed line).
In such a graphene junction the photovoltage $V_\mathrm{ph}$ is generated by the photo-thermoelectric effect:\cite{Gabor2011,Hsu2015,Lemme2011,Xu2010,Badioli2014}

\begin{equation}
V_\mathrm{ph} = \Delta S \Delta T
\end{equation}

where, $\Delta S = S_1 - S_2$ is the difference between the Seebeck coefficients of graphene on the left and right side of the junction and $\Delta T$ is the difference in electronic temperature at the junction and at the source/drain contacts.
The photo-thermoelectric effect dominates over other possible mechanisms of photovoltage generation due to the high Seebeck coefficient of graphene ($S \sim \SI{100}{\micro\volt\per\kelvin}$), which is in-situ tunable by gating.\cite{Gabor2011,Zuev2009}
By controlling the gate voltage on the two sides of the junction individually and recording the photocurrent we measure a 6-fold pattern, a clear sign that the photocurrent in our device is governed by the photo-thermoelectric effect\cite{Gabor2011,Hsu2015,Lemme2011,Xu2010,Badioli2014} (see supplement).
The highest responsivity measured when changing the gate voltages is obtained for $V_\mathrm{g1}=1.2$\,V and $V_\mathrm{g2}=-0.21$\,V.
This corresponds to a pn-configuration with a fairly low doping level of about 0.06\,eV (see supplement). 

The spectral responsivity (Fig.~\ref{fig2}a) is obtained by recording the photovoltage while tuning the wavelength of the quantum cascade laser source from 1000 to 1610\,cm$^{-1}$.
A strong photocurrent enhancement for the polarization perpendicular to the gap, peaking at 1515\,cm$^{-1}$, is observed.  
The peak around 1100\,cm$^{-1}$ is related with the SiO$_2$ surface phonon of the underlying substrate.\citep{Badioli2014}

In order to understand the observed behavior we use finite difference time domain (FDTD) simulations to model the scattering process of far-field light into HPPs and the subsequent HPP waveguiding and absorption of the HPPs in the graphene (solid lines in Fig.~\ref{fig2}a).
A good match with the experimentally observed spectral response is obtained (points in Fig.~\ref{fig2}a).
The simulated absorption spectrum shows a peak inside the reststrahlen band of h-BN.
The spatial distribution of the electric field inside the h-BN layers is shown in Fig.~\ref{fig2}d.
It is dominated by four rays which are launched at the edges of the split gate and undergo multiple reflections from the top and bottom surfaces.
The rays maintain a fixed angle with the $c$-axis.
This angle is related to the anisotropy of the permittivity via the analytical formula 
$\tan\theta(\omega)=i\sqrt{\epsilon_\mathrm{x,y}(\omega)}/\sqrt{\epsilon_\mathrm{z}(\omega)}$.\cite{Caldwell2014,Dai2015}
It predicts that $|\theta|$ changes from $\pi/2$ to $0$ as $\omega$ varies across the reststrahlen band. 
The unusual ray pattern of HPP emission in turn affects the spatial absorption pattern in graphene, which is shown in the simulated spectral-spatial pattern of Fig.~\ref{fig2}b.
This pattern is dominated by the four families of ``hot spots'' that correspond to the four HPP rays seen in Fig.~\ref{fig2}d.
The separation of hot spots within each family is $2d |\tan \theta|$, where $d = d_\mathrm{t} + d_\mathrm{b}$ is the total thickness of the h-BN layers (Fig.~\ref{fig1}c).

To investigate further the origin of the observed spectral peaks in the photocurrent we carried out scanning near-field photocurrent mapping of our devices.\cite{Woessner2016,Lundeberg2016}
In this technique a metallized atomic force microscopy tip is illuminated with an infrared laser and a near-field is generated at the apex of the tip.
This enables us to measure the photocurrent with a spatial resolution greatly exceeding the diffraction limit of light.
The representative results are shown in Fig.~\ref{fig2}c.
The device region measured includes the gap of the split gate and one graphene edge localized at the top of the frame.
The obtained photocurrent map reveals two series of sinusoidal spatial oscillations (fringes) rather than sharply peaked hot spots seen in Fig.~\ref{fig2}b.
These smooth oscillations can be explained if we recall that in an h-BN slab of small enough thickness $d$ the HPPs are quantized into discrete eigenmodes with in-plane momenta $k_l = \tan\theta (\pi l + \phi) / d$ where $l = 0, 1, 2, \ldots$ is the mode index and $\phi\sim 1$ is a phase shift that depends on the boundary conditions (see Fig.~\ref{fig3}b and e.g., Ref.~\citenum{Wu2015}).
The collimated rays seen in Fig.~\ref{fig2}d can be understood as coherent superpositions of many such modes emitted by the split-gate.
On the other hand, in the photocurrent microscopy the role of the HPP emitter is played by an AFM tip, which apparently couples predominantly to the $l  = 0$ mode.\cite{Dai2015a}
The horizontal fringes in Fig.~\ref{fig2}c are due to interference of $l = 0$ polariton waves launched by the tip, which is backreflected at the graphene edge leading to a fringe spacing corresponding to half the wavelength $\lambda_\mathrm{p} = 2\pi / k_0$ of this mode.\cite{Dai2015a}
The vertical fringes are due to interference of the $l=0$ partial wave launched at the split gate\cite{Yoxall2015} with the tip launched waves.
In this case the fringe spacing is $\lambda_\mathrm{p}$.
We do not observe HPPs launched by the tip and reflected by the gap as there are no vertical fringes with half the wavelength visible.
This interpretation enables us to extract $\lambda_p$ from the fringe spacing in the photocurrent maps.
For example, at 1428\,cm$^{-1}$ is $\lambda_\mathrm{p} = 460\pm 5$\,nm, which agrees with the calculated wavelength of 455\,nm.
The observed fringes parallel to the gap on the left of Fig.~\ref{fig2}c confirm that phonons are indeed launched by the split gate and are converted into photocurrent.

In order to better understand which parameters determine the absorption spectrum we also modelled the system  analytically.\cite{Wu2015}
In this model we approximate the electric fields at the bottom surface of the h-BN by the electric field inside the gap $-a < x < a$ cut along the $y$-axis in a perfectly conducting plane $z = 0$ in vacuum:

\begin{equation}
\label{eqn:E_x}
E_x(x,z=0) = \frac{V_0}{\pi}\, \mathrm{Re}\,  \frac{1}{\sqrt{a^2 - x^2}}\,. 
\end{equation}


Here $V_0$ is the voltage across the gap, which is proportional to the field of the incident beam (see inset Fig.~\ref{fig3}a).

The Fourier transform of $E_x$ is given by (Fig.~\ref{fig3}a)

\begin{equation}
\label{eqn:ft}
\widetilde{E}_x(k)=V_0 J_0 (ka),
\end{equation}

where $J_0(z)$ is the Bessel function of the first kind.

We then compute the field inside the h-BN-graphene layered structure using the transfer matrix method (Fig.~\ref{fig3}b) by assuming that Eq.~\ref{eqn:ft} represents the field incident on the structure from the bottom.
The assumption is not strictly self-consistent because it does not account for the backreaction of h-BN on the split-gate, in the form of the HPP rays reflected back to $z = 0$ plane.
A more accurate but also more complicated model that obeys the self-consistency condition is presented in Supplementary.
Unfortunately, that latter model can no longer can be solved in a closed form.
This is why here we use the simplified analytical model to illustrate the main features of the studied phenomena.
We calculate the Fourier transform of the in-plane electric field at the graphene surface as a function of momentum $k$.
The power absorbed in the graphene is then expressed as (Fig.~\ref{fig3}d):

\begin{equation}
\label{eqn_4}
p=\frac{1}{4\pi}\, \Re \sigma(\omega)\int|\widetilde{E}_x(k)|^2dk\,,
\end{equation}

where $\sigma(\omega)$ is the sheet conductivity of graphene at the laser frequency $\omega$.
Here, for simplicity, we neglect the spatial variation of $\sigma$ near the pn-junction as the hot spots responsible for the absorption are typically found some distance away from the junction (Fig.~\ref{fig3}c).

From this model description it becomes clear that the characteristic momentum $k \sim 1 / a$ provided by the junction plays a crucial role for the frequency of maximum absorption.
By calculating the inverse Fourier transform of $\widetilde{E}_x(k)$ we are able to also calculate the spatial profile of the electric field $E_x(x)$ and thus the spatial absorption profile (Fig.~\ref{fig3}c).
The validity of our analytic model can be seen by the close resemblance between the analytically calculated and FDTD simulated frequency dependent absorption profile (compare Fig.~\ref{fig2}b and~\ref{fig3}c). 

From this model, the origin of the peak in the spectral photoresponse is the competition between the following two processes: the dielectric losses in the h-BN and the (finite) momentum provided by the junction.
First, the losses in the h-BN contribute mainly to the low frequency side due to the imaginary part of the permittivity which peaks at the TO phonon frequency (1360\,cm$^{-1}$).
The impact of this effect on the device responsivity is enhanced by the obtuse angle with which the HPPs are launched, as the intensity of the HPPs reaching the graphene becomes smaller with travelled distance. Second, the momentum provided by the junction is responsible for the responsivity decay on the high frequency side.
Interestingly both of these effects depend on the h-BN thickness and on the gap size.
It is important to note that the impact of the h-BN thickness is twofold since it is also changing the HPPs dispersion.\citep{Caldwell2014}
Thus by choosing the geometrical parameters of the device, the device thickness and gap width, it is possible to tune the frequency as well as amplitude of the photocurrent maximum within the reststrahlen band of h-BN. 

In order to show this tunability, and to validate the physical model, we fabricated different device geometries.
Experimental responsivity spectra of the different devices are plotted in Fig.~\ref{fig4}a.
All the spectra were measured using the gate voltage configuration exhibiting the highest responsivity for the respective device.
They exhibit different peak frequencies and responsivities and the trend is well captured by the analytically calculated absorption spectra presented in Fig.~\ref{fig4}c.
The peak frequencies are plotted in Fig.~\ref{fig4}b as a function of the relevant geometrical parameters of the system.
These are the stack thickness $d=d_\mathrm{t}+d_\mathrm{b}$, where $d_\mathrm{t}$ and $d_\mathrm{b}$ are the bottom and top h-BN thicknesses (Fig.~\ref{fig1}c), and the split gate gap width $2a$.
The tunability of the investigated devices spans over 60\,cm$^{-1}$ and the peak frequencies obtained using both the FDTD simulations and the analytic model match the experimental ones.
In Fig.~\ref{fig4}d we plot the responsivities of the measured devices normalized to the highest one as a function of the peak frequencies.
We find that the responsivity follows a bell shaped curve (Fig.~\ref{fig4}d) suggesting that the optimal geometry would lead to a peak frequency where there is a trade off between low losses and high launching efficiency.
Using the analytic model we obtain a theoretical dependence of the frequency and of the absorbed power as a function of the stack thickness and the gap size (see supplement). 

In this simple analytic model the frequency dependence of the gap voltage $V_0$ (eq.~\ref{eqn:E_x}) is neglected. 
Thus, the coupling between the far-field light and the split gate is not taken fully into account.\cite{Alu2008}
This leads to some discrepancy between simple theory and experiment.
However, the mentioned above more sophisticated model based on the Galerkin method (see supplement) is in much better agreement with the experimental results and FDTD simulations (Fig.~\ref{fig4}b).

Finally, we will address the photodetection device performance.
We remark that the device operates at zero bias, leading to an extremely low noise level ($\sim\,4\,\mathrm{nV}/\sqrt{\mathrm{Hz}}$) from which we estimate a noise equivalent power (NEP) of $26\,\mathrm{pW}/\sqrt{\mathrm{Hz}}$ (see methods). 
From our simulations we found that the active area is about $\SI{2.5}{\micro\metre}^2$, i.e., only 2.5\% of the device area.
Thus, the device can be easily scaled to smaller dimensions, with the potential to enhance the performance by another factor of 40 because the total device resistance would be decreased and thus the Johnson-Nyquist noise would decrease as well leading to a lower NEP.
Current state-of-the art detectors based on other technologies are described in refs.~\citenum{Rogalski2011,Hamamatsu2011}.
At room temperature typically silicon bolometers are used. 
Our detectors can be further optimized to have similar detectivity as silicon bolometers, but offer several distinct advantages: it allows a smaller pixel size, higher operation speed and simpler fabrication as no suspension of the device is necessary.

Our novel nano-optoelectronic infrared detectors operate at room temperature, are highly efficient, and can be used for a wide range of on-chip sensing applications.

\section*{\textsf{Methods}}
{\small
\
\subsubsection*{\textsf{Sample fabrication}}
All the stack elements (top and bottom h-BN and graphene) are mechanically cleaved and exfoliated onto freshly cleaned Si/SiO$_2$ substrates. 
First the selected top h-BN is detached from the substrate using a PPC (poly-propylene carbonate) film and is then used to lift by Van der Waals forces the graphene and the bottom h-BN consecutively. 
The as-completed stack is released onto the split gate. 
The split gate electrodes are prepared by lithography, titanium (5\,nm)/gold (30\,nm) evaporation and focus ion beam irradiation to create the gap. 
The source and drain electrodes mask is designed in a AZ-5214 photoresist film by laser lithography and is exposed to a plasma of CHF$_3$/O$_2$ gases to partially etch the stack.
The graphene is finally contacted by the edges by evaporating titanium (2\,nm)/gold (30\,nm) and lift off in acetone. 
The recipe used for making those contacts is detailed in ref.~\citenum{Wang2013}. 

\subsubsection*{\textsf{Measurements}}
The device is illuminated by a linearly polarized quantum cascade laser with a frequency tunable from 1000 to 1610 cm$^{-1}$. 
The device position is scanned using a motorized xyz-stage. 
The laser is modulated at 128\,Hz using a chopper and the current at the junction is measured using a current pre-amplifier and lock-in amplifier.
The polarization of the light is controlled using a ZnSe wire grid polarizer. 
The light is focused using ZnSe lenses with a numerical aperture of $\sim$ 0.5. 
The power for each frequency is measured using a thermal power meter and the photocurrent spectra are normalized by this power to calculate the responsivity.

\subsubsection*{\textsf{Noise equivalent power (NEP) estimation}}
We calculate a NEP given by $NEP=S_\mathrm{noise}/R_\mathrm{internal} = 26\,\mathrm{pW}/\sqrt{\mathrm{Hz}}$  where $S_\mathrm{noise}$ is the voltage noise and $R_\mathrm{internal}$ is the internal responsivity. 
Because graphene pn-junction photodetectors operate at zero bias the electrical noise is of thermal Johnson-Nyquist type given by $S_\mathrm{noise}=\sqrt{4k_\mathrm{B} TR}=4\,\mathrm{nV}/\sqrt{\mathrm{Hz}}$. 
Where $k_\mathrm{B}$ is the Boltzmann constant, $T = 300$\,K and $R = 1\,\mathrm{k\Omega}$ is the resistance for which the calculated NEP is minimum corresponding to a carrier concentration of $n = 0.2\times 10^{12}\,\mathrm{cm}^{-2}$. 
The internal responsivity is given by  $R_\mathrm{internal}=R_\mathrm{external}/\eta=150$\,V/W where $R_\mathrm{external}=1$\,V/W is the experimental responsivity and $\eta=A_\mathrm{abs}/A_\mathrm{spot} =0.5$\%  the percentage of absorbed light. 
$A_\mathrm{spot}=\SI{491}{\micro\metre}^2$ is the laser spot area and $A_\mathrm{abs}=\sigma W=\SI{2.5}{\micro\metre}^2$ is the active area where $\sigma = \SI{250}{\nano\metre}$ is the absorption cross section obtained by FDTD simulations and $W=\SI{10}{\micro\metre}$ is the width of the device.

\subsubsection*{\textsf{Simulations}}
The full wave simulations were performed using Lumerical FDTD. 
The frequency dependent permittivity of the h-BN was taken from ref.~\citenum{Caldwell2014}. 
The optical conductivity of the graphene was calculated using the local random phase approximation at T = 300\,K with a scattering time of 500\,fs.
For each device the appropriate Fermi energy was simulated (see Supplement), however this did not influence the results significantly. 
In the simulations the Fermi energy of the graphene is spatially constant (see the comment after eq.~\ref{eqn_4}) but frequency dependent. 
A plane wave source was used and the absorption cross section was calculated by normalizing to the incident power. 
For simplicity the calculated absorption does not take into account the cooling length of the graphene nor the carrier density profile.
} 


\section*{\textsf{Acknowledgements}}
{\small
It is a great pleasure to thank Klaas-Jan Tielrooij for many fruitful discussions.
This work used open source software (www.matplotlib.org, www.python.org, www.povray.org). 
F.H.L.K. acknowledges financial support from the Spanish Ministry of Economy and Competitiveness,
through the “Severo Ochoa” Programme for Centres of Excellence in R\&D (SEV-2015-0522), 
support by Fundacio Cellex Barcelona,
the Mineco grants Ramón y Cajal (RYC-2012-12281) and Plan Nacional (FIS2013-47161-P and FIS2014-59639-JIN),
and support from the Government of Catalonia trough the SGR grant (2014-SGR-1535).
Furthermore, the research leading to these results has received funding from the European Union Seventh
Framework Programme under grant agreement no.696656 Graphene Flagship, and the ERC starting grant (307806, CarbonLight).
Y.G. and J.H. acknowledge support from the US Office of Naval Research N00014-13-1-0662.
P.A.-G. acknowledges funding from the Spanish Ministry of Economy and Competitiveness through the national projects FIS2014-60195-JIN.
%
\section*{\textsf{Author contributions}}
{\small
A.W. and F.H.L.K. conceived the experiment.
A.W. and R.P. performed the far-field experiments, analysed the data and wrote the manuscript. 
A.W. performed the FDTD simulations. 
A.W. and P.A.-G. performed the near-field experiments. 
D.D. and Y.G. fabricated the devices. 
M.B.L. and S.N. assisted with measurements, interpretation and discussion of the results. 
J.-S.W. and M.M.F. developed the analytic model. 
K.W. and T.T. synthesized the h-BN. 
All authors contributed to the scientific discussion and manuscript revisions.
}

\section*{\textsf{Competing Financial Interests}}
{\small
R.H. is co-founder of Neaspec GmbH, a company producing scattering-type scanning near-field optical microscope systems such as the ones used in this study. All other authors declare no competing financial interests.           
}

\newpage

\renewcommand\refname{Supplementary references}
\def\bibsection{\section*{\refname}} 


\setcounter{equation}{0}
\setcounter{figure}{0}
\setcounter{table}{0}
\setcounter{page}{1}
\makeatletter
\renewcommand{\theequation}{s\arabic{equation}}
\renewcommand{\thefigure}{S\arabic{figure}}
\renewcommand{\thetable}{S\arabic{table}}
\renewcommand{\thepage}{\roman{page}}
\renewcommand{\bibnumfmt}[1]{[S#1]}
\renewcommand{\citenumfont}[1]{S#1}


\pagebreak
\widetext
\begin{center}
\textbf{\Large Supplementary Material: \papertitle}
\end{center}

\section{Experiment}
\subsection{Electrical device characterization}
Electrical properties of the pn junction devices are first characterized by recording the drain current under a bias voltage of 5\,mV by sweeping simultaneously the two gate voltages ($V_\mathrm{g1}$, $V_\mathrm{g2}$) from $-3$ to 3\,V.
From this measurement we obtain the gate dependence of the device resistance when the entire graphene sheet is uniformly doped.
The experimental curve is fitted using the following equation:

\begin{equation}
R_\mathrm{tot} = R_\mathrm{c} + \frac{1}{e\mu\sqrt{n_0^2+(\epsilon\epsilon_0V_\mathrm{g}/e)^2}}
\end{equation}

where $R_\mathrm{c}$ is the sum of both contact resistances, $e$ is the elementary charge, $\mu$ is the carrier mobility,  $\epsilon_0$ is the vacuum permittivity, $\epsilon$ is the DC permittivity of h-BN, $n_0$ is the residual doping at the charge neutrality voltage and $V_\mathrm{g}$ is the gate voltage shift with respect to the charge neutrality voltage.
In the case of the device in Fig.~1 of the main text the fit leads to $R_\mathrm{c} = 1500\,\Omega\si{\micro\metre}$ and $\mu = 30963\,\si{\centi\metre\squared\per\volt\per\second}$.
The gate dependence map of the device resistance shown in Fig.~\ref{figS1}a is measured by sweeping both gates independently in the range ($-3$\,V, 3\,V).
The cross pattern is a clear sign of independent and stable gate efficiency and allows the access to the four doping configurations: pn, np, pp’ and nn’. 

\begin{figure*}[h]
\centering
\includegraphics{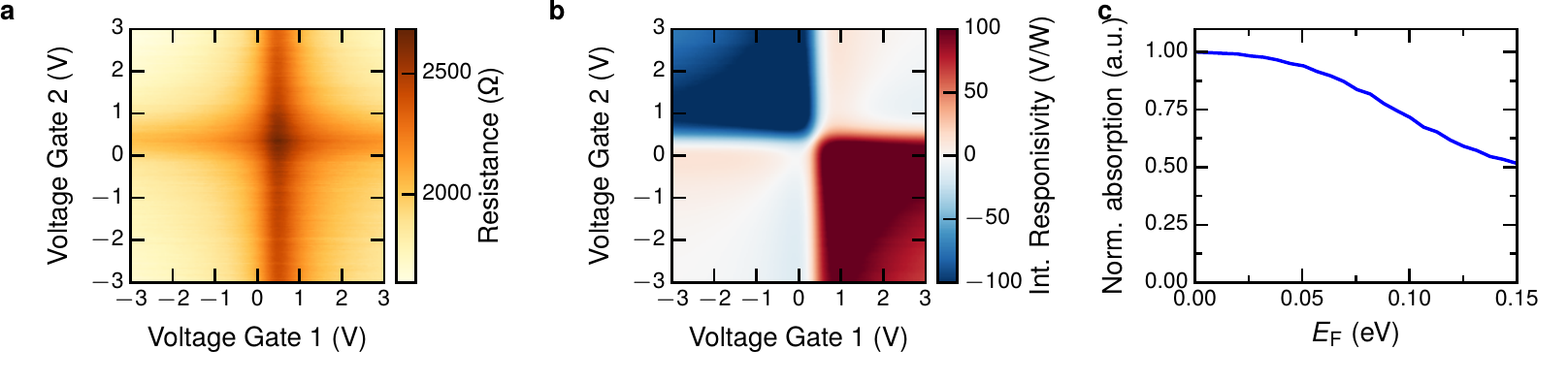} %
\caption{
\label{figS1}
\textbf{Gate dependence of resistance and photocurrent and Fermi energy dependent absorption.}
\textbf{a} Gate dependence of the device resistance.
\textbf{b} Gate dependence of the internal responsivity at laser frequency 1515\,cm$^{-1}$.
\textbf{c} Simulated normalized absorption at 1515\,cm$^{-1}$ for $V_\mathrm{g1} = -V_\mathrm{g2}$.
}
\end{figure*}

\subsection{Photocurrent generation}
The photoresponse is probed by focusing the laser beam on the junction and by sweeping the two gate voltages from $-3$\,V to 3\,V without biasing the device.
The obtained gate dependence of the responsivity exhibits a 6 fold pattern, a signature of the photo-thermoelectric effect in the photovoltage generation mechanism (Fig.~\ref{figS1}b).
A maximal internal responsivity of about 150\,V/W is measured in both pn and np configurations respectively at 1515\,cm$^{-1}$ for ($V_\mathrm{g1}=1.2$\,V, $V_\mathrm{g2}=-0.21$\,V) and ($V_\mathrm{g1}=-0.06$\,V, $V_\mathrm{g2}=1.26$\,V) which correspond to a fairly low carrier concentration of about $0.29\times10^{12}\,\mathrm{cm}^{-2}$ on one side of the junction and $0.23\times10^{12}\,\mathrm{cm}^{-2}$ on the other side.
At these optimal doping levels 0.112\,eV < 2$E_\mathrm{F}$ < 0.126\,eV is always lower than the energy range of the reststrahlen band (0.168\,eV < $E_\mathrm{L}$ < 0.198\,eV) meaning that the HPPs absorption by the graphene is never limited by Pauli blocking.
In order to be quantitative we simulated the graphene absorption at 1515\,cm$^{-1}$ (see main text) for a symmetric doping in the two regions of the junction variable in the range (0 < $E_\mathrm{F}$ < 0.15\,eV).
As shown on Fig.~\ref{figS1}c the absorption only drops by 10\% at the doping where the thermoelectric effect is the most efficient and by 50\% for the highest doping level explored ($E_\mathrm{F}= 0.136$\,eV).

\subsection{Device performance}
In Fig.~\ref{figS2}a and b we present respectively the gate dependence of the voltage noise and of the logarithm of the noise equivalent power ($\log(\mathrm{NEP})$).
The voltage noise of Johnson-Nyquist type is extracted from the gate dependence resistance map using:
\begin{equation}
S_\mathrm{noise}=\sqrt{4k_\mathrm{B} TR}
\end{equation}
The gate dependence of the NEP is calculated using:
\begin{equation}
\mathrm{NEP}=S_\mathrm{noise}/R_\mathrm{internal} 
\end{equation}
Here $S_\mathrm{noise}$ is the voltage noise and $R_\mathrm{internal}$ is the internal responsivity. 
A minimal value of $\mathrm{NEP}= 26\,$pW/$\sqrt{\mathrm{Hz}}$ is obtained for $V_\mathrm{g1}=1.35$\,V and $V_\mathrm{g2}=-0.48$\,V, a slightly different gate configuration than for the maximal responsivity. 
We have also measured the device time response $\tau$ using the quantum cascade laser (Block Engineering LaserScope) as a pulsed light source.
In the experiment we record simultaneously the beam reflection on the sample with a MCT detector and the photocurrent of the device.
Using a fast oscilloscope (Teledyne Lecroy HDO6104 1GHz High Definition Oscilloscope) to measure the MCT's output we get the laser pulse width $\tau_\mathrm{L}=\SI{0.24}{\micro\second}$ (Fig.~\ref{figS2}c).
The photoresponse of the device is amplified with a current amplifier (Femto DLPCA-200) and measured with the oscilloscope.
In Fig.~\ref{figS2}c we plot two line traces of the photocurrent obtained using two current amplifiers with two different cutoff frequencies of 200 kHz and 500 kHz.
These results clearly show that we are limited by the cutoff frequency of the amplifier thus we only get an upper value of the time response $\tau=\SI{2}{\micro\second}$.

\begin{figure*}[h]
\centering
\includegraphics{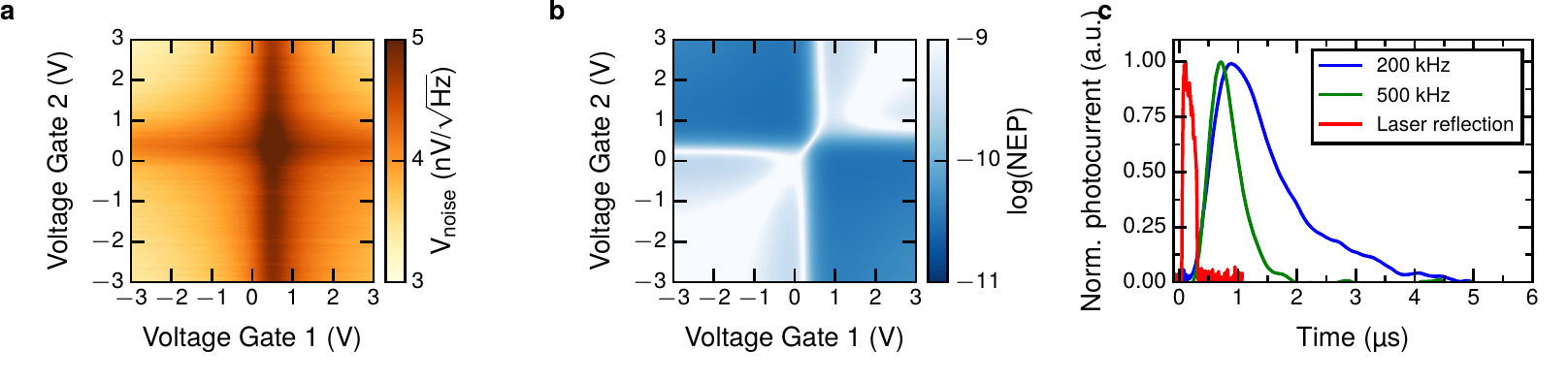} %
\caption{
\label{figS2}
\textbf{Gate dependence of resistance and NEP and photocurrent response time.}
\textbf{a} Gate dependence of the Johnson Nyquist noise.
\textbf{b} Gate dependence of the $\log(\mathrm{NEP})$.
\textbf{c} Measurement of the device time response. 
Laser pulse reflection (red).
Photocurrent measured using a current amplifier with 500\,kHz cutoff frequency (green). 
Photocurrent measured using a current amplifier with 200\,kHz cutoff frequency (blue). 
}
\end{figure*}

\subsection{Graphene absorption}
In Fig.~\ref{figS3}b we show the frequency dependent absorption profile of the HPPs by the graphene  $\sigma(x)|E(x)|^2$ simulated using the analytic model.
The intensity of the electric field $|E(x)|^2$ and the optical conductivity of the graphene pn junction $\sigma(x)$ in the case of symmetric doping ($E_\mathrm{F} = \pm 0.1$\,eV) are respectively presented in Fig.~\ref{figS3}a and~\ref{figS3}c.
This moderate doping level is the onset of the Pauli blocking, thus the absorption is slightly reduced in the n and p region of the junction but remains unchanged in the intrinsic part of the junction.
In Fig.~\ref{figS3}d we compare the absorbed power spectra of uniformly doped graphene ($E_\mathrm{F} = 0.1$\,eV) and of graphene pn junction ($E_\mathrm{F} = \pm 0.1$\,eV).
We observe an extremely weak shift toward the high frequency in the case of the junction.
This is explained by the higher weight of the HPPs absorption in the intrinsic part of the junction which takes place at higher frequency.    

\begin{figure*}[h]
\centering
\includegraphics{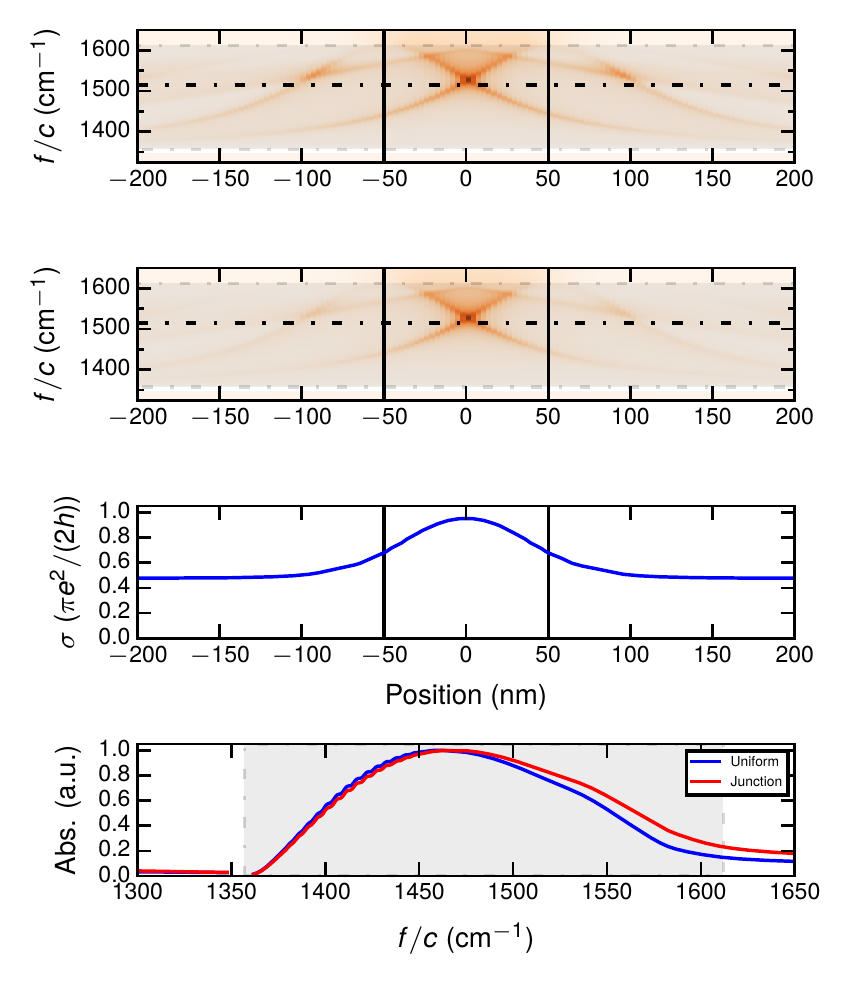} %
\caption{
\label{figS3}
\textbf{Position dependent absorption in the graphene.}
\textbf{a} Frequency dependent electric field profile $|E(x)|^2$ simulated with the analytic model.
\textbf{b} Simulated frequency dependent absorption profile $\sigma(x)|E(x)|^2$. 
The dashed dotted line indicates where the maximum photocurrent responsivity is observed experimentally. 
\textbf{c} Normalized optical conductivity of the graphene $\sigma(x)$.
The doping level is symmetric in the p and n region $E_\mathrm{F} = \pm 0.1$\,eV. 
\textbf{d} Simulated spectra of the absorbed power in the case of uniformly doped graphene ($E_\mathrm{F} = 0.1$\,eV) (blue) and in the case of the pn junction (symmetric doping $E_\mathrm{F} = \pm 0.1$\,eV).
}
\end{figure*}

\subsection{Photoresponse tunability}
Our analytic model has revealed the effect of two relevant geometrical parameters, the stack thickness ($d_\mathrm{t}+d_\mathrm{b}$) and the split gate gap width ($2a$), on both the frequency and the maximum absorbed power.
In Fig.~\ref{figS4} we show these dependences for a large set of geometries: 10\,nm < $d_\mathrm{t}+d_\mathrm{b}$ < 150\,nm and 30\,nm < $2a$ < 150\,nm. 
The frequency dependence has the shape of a sloping plane of equation:
\begin{equation}
f/c(d_\mathrm{t}+d_\mathrm{b},2a)=\alpha(d_\mathrm{t}+d_\mathrm{b})+\beta 2a+1405
\end{equation}
Where $\alpha=0.428\times10^9$ and $\beta-0.292\times 10^9$.

The map of the absorbed power reveals that the optimal geometry is when both the gap width and sample thickness are small.

\begin{figure*}[h]
\centering
\includegraphics{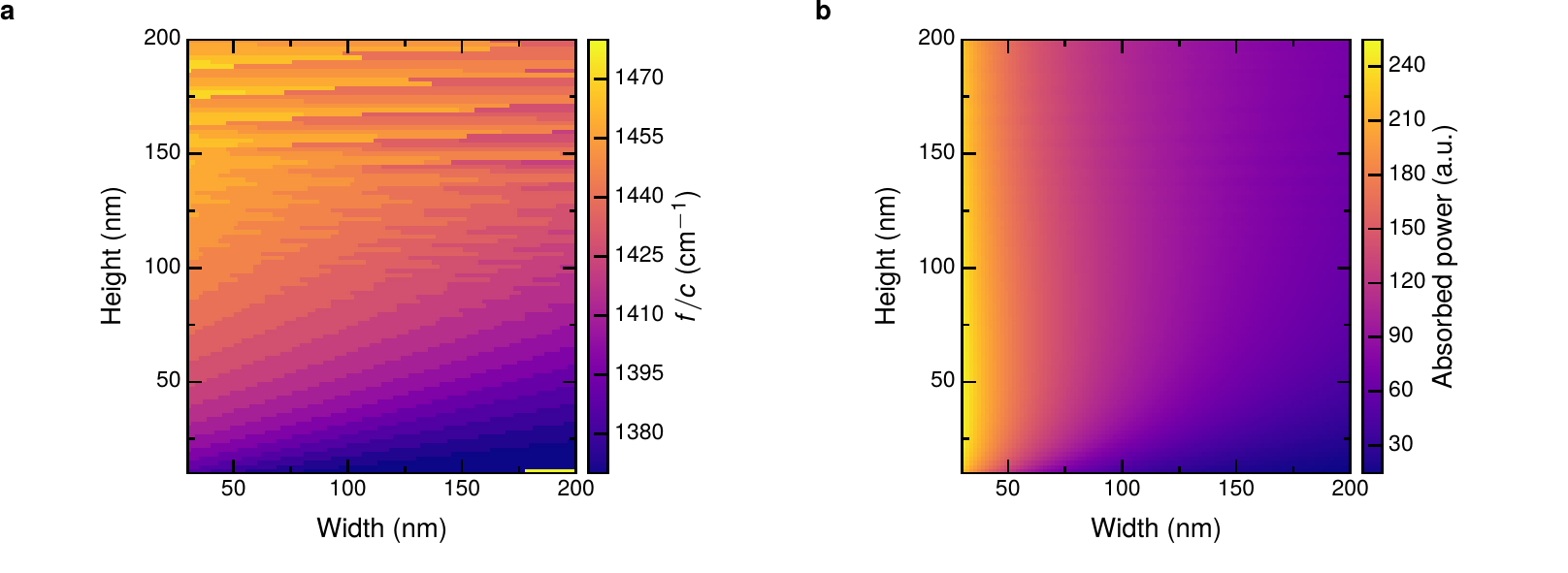} %
\caption{
\label{figS4}
\textbf{Analytic model.}
\textbf{a} Geometrical dependence of the peak frequency from the analytical model.
\textbf{b} Geometrical dependence of the absorbed power from the analytical model. 
}
\end{figure*}

\subsection{Device parameters}


%
\begin{table}[h!]
\begin{center}
\begin{tabular}{crcrcccc}
\hline\hline\\[-0.1in]
No. & $d_t\,(\mathrm{nm})$ & $d_b\,(\mathrm{nm})$
    & $2a\,(\mathrm{nm})$ & $d_{\mathrm{gate}}\,(\mathrm{nm})$
    & $\omega_{\mathrm{exp}}\,(\mathrm{cm}^{-1})$
    & $\omega_{\mathrm{sim}}\,(\mathrm{cm}^{-1})$
    & $E_F\,(\mathrm{meV})$\\[0.5ex]
\hline\\[-0.1in]
1 & 3\phantom{10}  & 30 & 70\phantom{00}  & 15 & 1490 & 1464 & 72\\
2 & 55\phantom{10} & 50 & 100\phantom{00} &	30 & 1520 & 1515 & 52\\
3 & 9\phantom{10}  & 27 & 150\phantom{00} & 15 & 1460 & 1447 & 37\\
4 & 17\phantom{10} & 60 & 60\phantom{00}  & 30 & 1512 & 1505 & 82\\
\hline\hline
\end{tabular}
\caption{Parameters of the experimental devices.
The first column is the device number.
The dimensions $d_t$, $d_b$, and $2a$
are indicated in Fig.~\ref{fig:G_hBN}.
Variable $d_{\mathrm{gate}}$ is the thickness of the Au split-gate. Frequencies $\omega_{\mathrm{exp}}$ and
$\omega_{\mathrm{sim}}$ are the positions of thermocurrent maxima in, respectively, experiment and numerical simulations.
$E_F$ is the Fermi energy of graphene.
}
\end{center}
\label{tbl:devices}
\end{table}

\section{Theory}
\subsection{The model geometry and the thermocurrent}
\label{sec:thermocurrent}

\begin{figure}[b]
\hspace{120pt}\includegraphics[width=10cm]{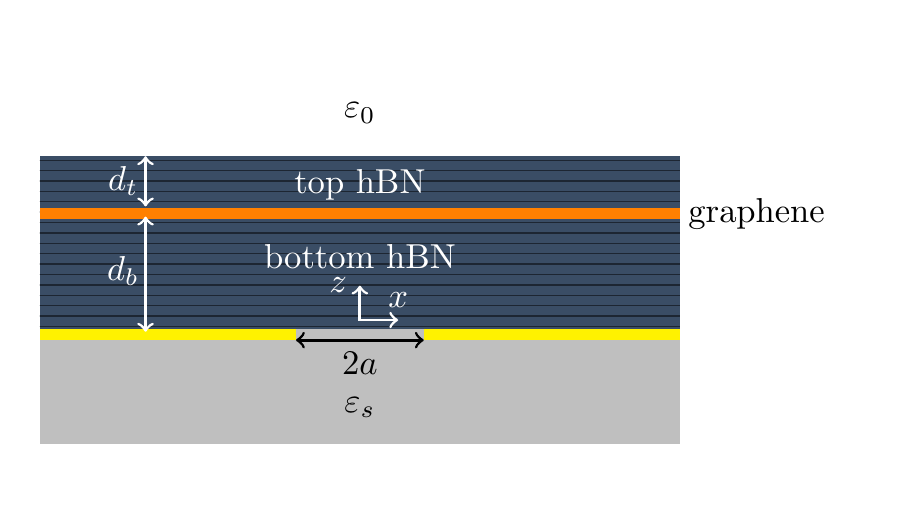}
\caption{Device schematics.}\label{fig:G_hBN}
\end{figure}

The geometry of the device under consideration is illustrated in Fig.~\ref{fig:G_hBN}.
It consists of a graphene layer sandwiched between two hBN slabs of thickness $d_t$ and $d_b$, a thin metallic film (split-gate) containing a gap of width $2a$, and a dielectric substrate (SiO$_2$). 
The coordinate system is shown in Fig.~\ref{fig:G_hBN}, where the middle of the gap is at $x=0$.
We assume that the sheet conductivity $\sigma(\omega, x)$ of graphene is a known function of frequency $\omega$ and position $x$.

In the experiment the device is illuminated by an infrared beam
that creates a distribution of
the electric field $E_g = E(z = d_b)$ in graphene
(here and below the factors $e^{-i\omega t}$ are omitted).
The corresponding Joule heating
\begin{equation}
p(x,\omega) = \Re \textrm{e}\, \sigma(\omega, x) |E_g(x,\omega)|^2
\label{eqn:p}
\end{equation}
causes an increase of the electron temperature $T(x)$ in graphene.
The gradient of $T(x)$ generates a dc thermocurrent $j$, which is measured.
The problem of computing the thermophotocurrent in a graphene $p$-$n$ junction has been addressed in prior literature~\cite{Gabor2011}. 
The change of temperature $\Delta T$ is shown to be proportional to $P(\omega)$ \cite{Gabor2011}, where
\begin{equation}
P(\omega)=\int_{-\infty}^{\infty}p(x,\omega)dx.
\end{equation}
Regardless of the proportional coefficient, the frequency dependence of the thermophotocurrent is controlled by the total Joule heating power $P(\omega)$.
The remainder of this note is devoted to computing this quantity.

\subsection{Split-gate as an antenna}
\label{sec:antenna}
The gold split-gate can be regarded as an antenna illuminated by the laser.  
The illumination induces a voltage difference $V_0$ across the gap. 
Because of retardation effects, the voltage $V_0$ is actually a function of frequency $\omega$, i.e., $V_0(\omega)$.
To obtain $V_0(\omega)$, we approximated the split-gate by a slotline made of a perfect metal.
The methods for electromagnetic problems of slotlines are well developed.\cite{Garg2013,Bouwkamp1954}
We follow the approach in Ref.~\cite{Popov2012}, which dealt almost the same geometry as in our case.

Suppose the system is illuminated by a $P$-polarized plane wave, that is, by an incident wave whose electric field is in the $x$--$z$ plane and whose magnetic field is along the $y$-axis.

The voltage $V_0$ is related to the electric field by  
\begin{equation}
V_0 (\omega)= \int\limits_{-a}^{a} d x E_s(x,\omega)\,,
\label{eqn:V_0_def}
\end{equation}
where $E_s(x) \equiv E^x(x, 0)$ is the field inside the slot ($-a<x<a$).
We denote vacuum and hBN as Medium~0 and Medium~1, respectively.
In  Medium $n$ the eigenmodes are plane waves with tangential momenta $q$ and $z$-axis momenta $\pm k^z_n(q,\omega)$ where
\begin{equation}
k^z_n(q,\omega) = \sqrt{\varepsilon^\bot_n(\omega) k_0^2 - \frac{\varepsilon^\bot_n(\omega)}{\varepsilon^z_n(\omega)} q^2}\,,
\qquad
\Im\mathrm{m}\, k^z_n \geq 0\,.
\label{eqn:k^z_n}
\end{equation}

We obtain the desired integral equation for the electric field
on the slot \cite{Popov2012}:
\begin{equation}
\boxed{
\quad \int\limits_{-a}^{a} d x' G(x - x,\omega) E_s(x')
 = -2 i  \ell_0(q_0) e^{i q_0 x}\,,
 \quad -a < x < a\,.\quad
}
\label{eqn:E_s_equation}
\end{equation}
where the Green function is defined by
\begin{equation}
G(x,\omega) = \int \frac{d q}{2\pi i} e^{i q x} \left[
             \ell_1(q,\omega)\,
             \frac{1 + r(q,\omega)}{1 - r(q,\omega)}
           + \ell_2(q,\omega) \right].
\label{eqn:G}
\end{equation}
and in the Fourier domain,
\begin{equation}
i\widetilde{G}(q,\omega)=  \left[
             \ell_1(q,\omega)\,
             \frac{1 + r(q,\omega)}{1 - r(q,\omega)}
           + \ell_2(q,\omega) \right].
\label{eqn:Gq}
\end{equation}
Function $r = r(q,\omega)$, which is given by
\begin{equation}
r = e^{i \alpha_b}
     \frac{r_{1 0} - r_g \left(2 r_{1 0}- e^{-i \alpha_t}\right)}
          {e^{-i \alpha_t} - r_g r_{1 0}}\,,\qquad
\alpha_{b, t} = 2 k_1^z d_{b, t}\,,
\label{eqn:r}
\end{equation}
represents the total reflection coefficient of a wave launched upward from the gate.
The coeffecients $r_{10}$ and $r_g$ describe the reflections due to the hBN-vacuum interface and graphene.
They are expressed as
\begin{equation}
r_{10}(q,\omega) = \frac{\ell_0 - \ell_1}{\ell_0 + \ell_1}\,,
\qquad \ell_n(q,\omega) \equiv \frac{\varepsilon^\bot_n}{k^z_n}\,.
\label{eqn:r_mn}\\
\end{equation}
\begin{equation}
r_{g}(q,\omega)
 = \frac{\frac{ 2\pi \sigma(\omega)}{\omega}}
        {\ell_1(q,\omega) + \frac{2\pi \sigma(\omega)}{\omega}}\,.
\label{eqn:r_g}
\end{equation}

 The field $E_g(x)$ in the graphene plane are given by
\begin{align}
E_{g}(x,\omega) &= \int \frac{d q}{2\pi} e^{i q x}\,
\frac{e^{i k_1^z d_b} - r(q,\omega) e^{-i k_1^z d_b}}
     {1 - r(q,\omega)}\,
     \widetilde{E}_s(q)\,,
\label{eqn:E_g3}
\end{align}
where the tilde denotes the Fourier transform in $x$.

\subsection{Variational method for the field distribution}
\label{sec:solution}
The electric field $E_g(x,\omega)$ can be obtained as a certain quadrature over the field $E_s(x,\omega)$ on the slot, as shown in Eq.~\eqref{eqn:E_g3}.
Thus, the brunt of the calculation is to compute $E_s(x,\omega)$.
The standard approach is the Galerkin method where one seeks $E_s(x,\omega)$ as a series of basis functions~\cite{Garg2013}
\begin{equation}
E_s(x,\omega) = \sum\limits_{k = 0}^{\infty} c_k(\omega) f_k(x)\,,
\qquad
f_k(x) = \frac{1}{\sqrt{a^2 - x^2}}\, T_k\left(\frac{x}{a}\right)\,,
\quad
-a < x < a\,,
\label{eqn:E_s_Chebyshev}
\end{equation}
where $T_l(z) = \cos(l \arccos z)$ is the Chebyshev polynomial of degree $l$.
The advantages of working with the Chebyshev polynomials are two-fold, the fast convergence and the closed-form representation of the field in the Fourier domain:
\begin{equation}
\widetilde{E}_s(q,\omega) = \pi 
\sum\limits_{k = 0}^{\infty} (-i)^k c_k(\omega) J_k(q a)\,,
\qquad
V_0 = \widetilde{E}_s(0) = \pi c_0(\omega)\,,
\label{eqn:E_s_q}
\end{equation}
where $J_\nu(z)$ is the Bessel function of the first kind.
The coefficients $c_{n}(\omega)$ is given by \cite{Popov2012}
\begin{equation}
	 \mathbf{c}(\omega) = -{2i}  \ell_0(q_0)\mathbf{M}^{-1} \mathbf{e}\,,
\label{eqn:matrix_eq}
\end{equation}
where the matrix $\mathbf{M}(\omega)$ has the elements
\begin{equation}
M_{mn}(\omega) = (-1)^{m - n}
\int\limits_0^{\infty} J_{2n}(qa) J_{2m}(qa)
\widetilde{G}(q,\omega) {d q}
\label{eqn:M_mn}
\end{equation}
and the column vector $\mathbf{c}(\omega)$ consists of the coefficients $c_{0}(\omega), c_{2}(\omega), \ldots$
The column vector $\mathbf{e}$ has a single nonzero entry
$e_0 = 1$ because Chebyshev polynomials $T_0(z) \equiv 1$ and $T_k(z)$,
$k \neq 0$, are orthogonal with the weight
$1 / \sqrt{1 - z^2}$.
In practice, in order to solve Eq.~\eqref{eqn:matrix_eq} we have to truncate $\mathbf{M}(\omega)$ to a finite order.
To find how large $N$ needs to be, we can
monitor the coefficients $c_i$ as $N$ is incremented until they converge to within the desired tolerance.
From our simulations we found that for computing the accurate field profile $E_s^x(x)$ as many as $N \sim 10$ basis functions may be necessary.
However, for the calculation of the gap voltage $N = 1$, i.e., the single-mode approximation (SMA) is typically sufficient.

Figure~\ref{fig:ab} shows the power due to the Joule heating for Device 1-4, given by the variational method. 
\begin{figure}
\begin{center}
\includegraphics{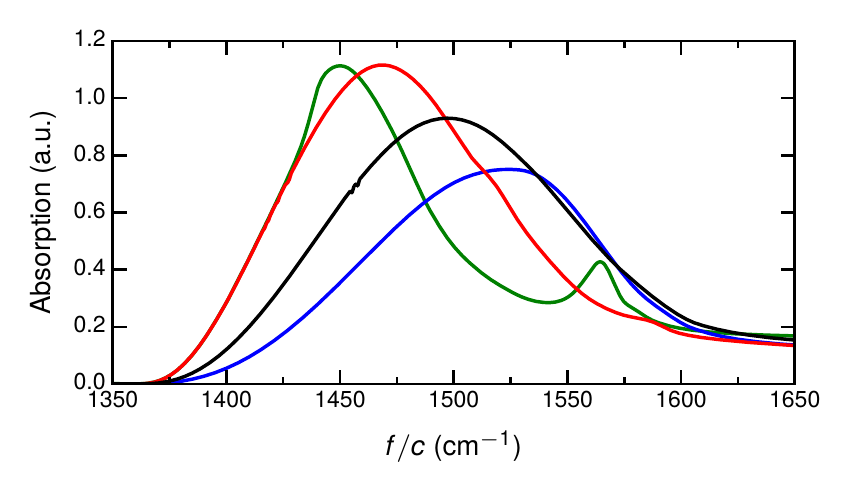}
\end{center}
\caption{Power due to the Joule heating calculated using the variational method for the field distribution outlined in this supplement.}\label{fig:ab}
\end{figure}


\end{document}